\documentclass[a4paper,twocolumn,showpacs,superscriptaddress,floatfix]{revtex4}
\usepackage{graphicx}
\usepackage{epsf}
\usepackage{epsfig}
\usepackage{amssymb,amsmath}
\usepackage[usenames]{color}
\usepackage{amssymb}
\usepackage{times}

\newcommand{\be}{\begin{equation}}
\newcommand{\ee}{\end{equation}}
\newcommand{\bea}{\begin{eqnarray}}
\newcommand{\eea}{\end{eqnarray}}
\newcommand{\nn}{\nonumber}

\font\tenscr=rsfs10 scaled1100
\font\sevenscr=rsfs7 
\font\fivescr=rsfs5 
\skewchar\tenscr='177
\skewchar\sevenscr='177
\skewchar\fivescr='177
\newfam\scrfam
\textfont\scrfam=\tenscr
\scriptfont\scrfam=\sevenscr
\scriptscriptfont\scrfam=\fivescr

\begin{document}

\title{Black hole Area-Angular momentum inequality in non-vacuum
spacetimes}

\author{Jos\'e Luis Jaramillo}
\affiliation{
Max-Planck-Institut f{\"u}r Gravitationsphysik, Albert Einstein
Institut, Am M\"uhlenberg 1 D-14476 Potsdam Germany 
}
\affiliation{
Laboratoire Univers et Th\'eories (LUTH), Observatoire de Paris, CNRS, 
Universit\'e Paris Diderot, 92190 Meudon, France
}

\author{Mart\'\i n Reiris}
\affiliation{
Max-Planck-Institut f{\"u}r Gravitationsphysik, Albert Einstein
Institut, Am M\"uhlenberg 1 D-14476 Potsdam Germany 
}

\author{Sergio Dain}
\affiliation{
Facultad de Matem\'atica, Astronom\'\i a y F\'\i sica, FaMAF, 
Universidad Nacional de C\'ordoba \\
Instituto de F\'\i sica Enrique Gaviola, IFEG, CONICET, \\
Ciudad Universitaria (5000) C\'ordoba, Argentina
}
\affiliation{
Max-Planck-Institut f{\"u}r Gravitationsphysik, Albert Einstein
Institut, Am M\"uhlenberg 1 D-14476 Potsdam Germany 
}
\begin{abstract}
We show that the area-angular momentum inequality $A\geq 8\pi|J|$ holds for 
axially symmetric closed outermost stably marginally trapped surfaces. 
These are horizon sections (in particular, apparent horizons) contained in 
otherwise generic  non-necessarily axisymmetric black hole spacetimes, 
with non-negative cosmological constant and
whose matter content satisfies the dominant energy condition. 
\end{abstract}


\maketitle

\emph{Introduction.~}
Isolated and stationary black holes cannot rotate arbitrarily fast. The total
angular momentum $J$ in Kerr solutions that are  consistent with cosmic censorship 
(i.e. without {\em naked} singularities)
is bounded from above by the square of the total mass $M$. The heuristic
{\em standard picture of gravitational collapse} \cite{Pen73} 
emphasizes the physical relevance of this bound and suggests its generic 
validity beyond idealized situations. The total mass-angular momentum inequality 
$J\leq M^2$ has been indeed  extended to the dynamical 
case of vacuum axisymmetric black hole spacetimes \cite{Dai06a,Dain:2005qj,Dain:2006wb,
Chrus07a,Chrus07b,Chrusciel:2009ki}. However, this inequality involves
global quantities. In order to gain further insight into the gravitational 
collapse process in the presence of matter and/or multiple horizons, it 
is also desirable to have a quasi-local version of the inequality.
This attempt encounters immediately, though, the ambiguities associated 
with the quasi-local definition
of gravitational mass and angular momentum. In this context, an alternative (but
related) bound on the angular momentum can be formulated in terms
of a horizon area-angular momentum inequality $A \geq 8\pi |J|$.
This inequality was 
conjectured for the non-vacuum axisymmetric stationary case (actually, 
including the charged case) with matter {\em surrounding} the horizon 
in \cite{AnsPfi07} and then proved in 
\cite{HenAnsCed08,Hennig:2008zy,Ansorg:2010ru}, whereas its validity
in the vacuum axisymmetric dynamical case was conjectured and
discussed in \cite{Dain:2010qr}, partial results were given in 
\cite{Acena:2010ws,GabachClement:2011kz}
 and a complete proof in \cite{Dain:2011pi}.
Equality holds in the extremal case. Here 
we reconciliate and extend both results by proving the validity of the
inequality in fully general dynamical non-vacuum spacetimes
(matter on the horizon allowed), only requiring
axisymmetry on the horizon. 
The rest of this letter is devoted 
to prove this result.

\emph{The dynamical non-vacuum case.~} Proofs of $A \geq 8\pi |J|$ require 
some kind of
geometric stability condition characterizing the surface ${\cal S}$ for 
which the inequality is proved.
On the one hand, in the non-vacuum stationary case discussed in 
\cite{HenAnsCed08,Hennig:2008zy,Ansorg:2010ru} surfaces ${\cal S}$ 
are taken to be 
sections of black hole horizons modeled as
outer trapping horizons \cite{Hay94}. This entails, first, the vanishing of the
expansion $\theta^{(\ell)}$ associated with light rays emitted from ${\cal S}$ 
along the (outgoing) null normal $\ell^a$ [i.e. ${\cal S}$
is a marginally trapped surface] and, second,
that when moving towards the interior of the black hole one finds 
fully trapped surfaces, so that the variation of $\theta^{(\ell)}$ 
along some future ingoing null normal $k^a$ is negative (outer condition):
$\delta_k \theta^{(\ell)} < 0$ (see \cite{Booth:2007wu} for a detailed
discussion of this condition in the context of black hole extremality). 
The latter inequality acts as a stability condition on ${\cal S}$
and, actually, is closely related to the {\em stably outermost} condition imposed 
on marginally trapped surfaces contained in spatial 3-slices $\Sigma$ when proving
the existence of dynamical trapping horizons \cite{AndMarSim05}. Such 
{\em stably outermost} condition
means that the variation of  $\theta^{(\ell)}$ along some outward deformation 
of $\cal S$ in the slice $\Sigma$ is non-negative. That is, 
$\delta_v \theta^{(\ell)} \geq 0$ for some spacelike {\em outgoing} vector 
$v$ tangent to $\Sigma$ (see the generalization to spacetime normal vectors in 
\cite{AndMarSim08}; we refer to  \cite{AndMarSim05} for a discussion
on operator $\delta_v$). 
Regarding now  the vacuum dynamical case 
in \cite{Dain:2011pi}, the inequality $A \geq 8\pi |J|$ is first proved
for stable minimal surfaces ${\cal S}$ in a spatial maximal slice $\Sigma$, 
i.e. ${\cal S}$ is a local
minimum of the area when considering arbitrary deformations of ${\cal S}$ in $\Sigma$,
and then generalized for arbitrary surfaces, in particular horizon sections.

The present discussion of the inequality $A \geq 8\pi |J|$ 
closely follows the strategy and steps in \cite{Dain:2011pi}, adapting them 
to the use of a stability condition in the spirit of those in
\cite{HenAnsCed08,Hennig:2008zy,AndMarSim05,AndMarSim08,Booth:2007wu}, i.e. based on 
marginally trapped surfaces rather than on minimal surfaces. 
In the line of \cite{AndMarSim05,AndMarSim08} we will refer to
a marginally trapped surface ${\cal S}$ as {\em (spacetime) stably outermost} 
(see Definition 1 below)
if for some outgoing space-like vector or outgoing past null vector $X^a$ 
it holds $\delta_X \theta^{(\ell)} \geq 0$. 
Then, it follows:

\vspace{0.2cm}
{\bf Theorem 1.~} {\em Given an axisymmetric closed marginally trapped
surface ${\cal S}$ satisfying the (axisymmetry-compatible)
spacetime stably outermost condition, in a spacetime  with non-negative cosmological constant
and fulfilling the dominant energy condition, 
it holds the inequality
\bea
\label{e:inequality}
A \geq 8\pi |J| \ \ ,
\eea
where $A$ and $J$ are the area
and gravitational 
(Komar) angular momentum of ${\cal S}$. If equality holds, then 
${\cal S}$ has the geometry of an extreme Kerr throat sphere and, in 
addition, if vector $X^a$
in the stability condition can be found to be spacelike then
${\cal S}$  is a section of a non-expanding horizon.
}

\vspace{0.3cm}
Note that axisymmetry is only required on the horizon surface
(this includes the intrinsic geometry of ${\cal S}$ and a certain 
component of its extrinsic geometry, see below), so that
$J$ accounts solely for the angular momentum of the black hole (horizon) 
in an otherwise generically non-axisymmetric spacetime.
Actually, no other geometric requirement is imposed outside ${\cal S}$.
Regarding the topology of the marginally 
trapped surface ${\cal S}$, this is always a topological
sphere (for $J\neq 0$) as a consequence of the stability condition combined
with the dominant energy condition. Therefore, we can assume in the following
that ${\cal S}$ is a sphere $S^2$ without loss of generality.

\emph{Elements in the proof.~} The proof in \cite{Dain:2011pi} has two parts.
First, a geometric part providing a lower bound on the area $A$.
And second, a part making use of variational principles to relate 
that lower area bound to an upper bound on the angular momentum $J$ 
and, in a subsequent step, to prove rigidity. Here we 
recast the first geometric part in the new setting and 
recover exactly the functional needed in the second variational part, so that
results in \cite{Acena:2010ws,Dain:2011pi} 
can be directly applied.

Let us first introduce some notation and 
consider a closed orientable 2-surface ${\cal S}$ embedded in a spacetime $M$
with metric $g_{ab}$ and Levi-Civita connection $\nabla_a$, 
satisfying the dominant energy condition and with non-negative cosmological
constant $\Lambda\geq 0$. We denote the induced metric on ${\cal S}$ 
as $q_{ab}$, with Levi-Civita
connection $D_a$, Ricci scalar ${}^2\!R$ and volume element $\epsilon_{ab}$ 
(we will
denote by $dS$ the area measure on ${\cal S}$). Let us consider null vectors
$\ell^a$ and $k^a$ spanning the normal plane to ${\cal S}$ and
normalized as $\ell^a k_a = -1$, leaving a (boost) rescaling freedom 
$\ell'^a =f \ell^a$, $k'^a = f^{-1} k^a$. 
The expansion $\theta^{(\ell)}$ and the shear 
$\sigma^{(\ell)}_{ab}$ associated with the null normal $\ell^a$ are given by
\bea
\label{e:expansion_shear}
\theta^{(\ell)}=q^{ab}\nabla_a\ell_b \ \ , \ \  
\sigma^{(\ell)}_{ab}=  {q^c}_a {q^d}_b \nabla_c \ell_d - \frac{1}{2}\theta^{(\ell)}q_{ab} \ ,
\eea
whereas the normal fundamental form $\Omega_a^{(\ell)}$ is 
\bea
\label{e:Omega}
\Omega^{(\ell)}_a = -k^c {q^d}_a \nabla_d \ell_c \ .
\eea
Transformation rules under a null normal rescaling are
\bea
\label{e:null_transformations}
\theta^{(\ell')}=f \theta^{(\ell)} \ , \ 
\sigma^{(\ell')}_{ab}= f\sigma^{(\ell)}_{ab} \ , \
\Omega^{(\ell')}_a = \Omega^{(\ell)}_a + D_a(\mathrm{ln}f).
\eea

\vspace{0.3cm}
We characterize now the surfaces ${\cal S}$ for which the result 
in Theorem 1 holds.
First, 
we impose ${\cal S}$ to be axisymmetric, with
axial Killing vector  $\eta^a$, i.e. ${\cal L}_\eta q_{ab}=0$. 
More precisely, $\eta^a$ 
defined on ${\cal S}$ has closed integral curves and
 vanishes exactly at two points. We normalize vector $\eta^a$
so that its integral curves have an affine length
of $2\pi$.
The associated  gravitational angular momentum (the Komar one, 
if $\eta^a$ can be extended as a  Killing vector to a neighbourhood of ${\cal S}$) 
is expressed in terms of $\Omega_a^{(\ell)}$ as
\bea
\label{e:angular_momentum}
J = \frac{1}{8\pi}\int_{\cal S} \Omega_a^{(\ell)} \eta^a dS \ ,
\eea
where the divergence-free character of $\eta^a$ together with 
the transformations 
properties of $\Omega_a^{(\ell)}$ in (\ref{e:null_transformations})
guarantee the invariance of $J$ under rescaling
of the null normals. 
We also assume a tetrad 
$(\xi^a, \eta^a, \ell^a, k^a)$ on ${\cal S}$, adapted to axisymmetry 
in the sense that ${\cal L}_\eta \ell^a = {\cal L}_\eta k^a = 0$,
with $\xi^a$ a unit 
vector tangent to ${\cal S}$ and orthogonal to $\eta^a$, 
i.e.  $\xi^a\eta_a=\xi^a\ell_a=\xi^ak_a=0$, $\xi^a\xi_a=1$.
We can then write the induced metric on ${\cal S}$ as
$q_{ab}=\frac{1}{\eta}\eta_a\eta_b + \xi_a\xi_b$,
with $\eta=\eta^a\eta_a$, so that
\bea
\label{e:Omega_eta_xi}
\Omega^{(\ell)}_a &=& \Omega^{(\eta)}_a + \Omega^{(\xi)}_a \nn  \\ 
\Omega^{(\ell)}_a{\Omega^{(\ell)}}^a &=& \Omega^{(\eta)}_a{\Omega^{(\eta)}}^a + 
\Omega^{(\xi)}_a{\Omega^{(\xi)}}^a \ ,
\eea
with $\Omega^{(\eta)}_a= \eta^b\Omega^{(\ell)}_b \eta_a/\eta$
and $\Omega^{(\xi)}_a= \xi^b\Omega^{(\ell)}_b \xi_a$. 
In addition, we demand $\Omega^{(\ell)}_a$ to be also axisymmetric, 
${\cal L}_\eta \Omega^{(\ell)}_a=0$. 
Second, ${\cal S}$ is taken to be a marginal trapped surface:
$\theta^{(\ell)}=0$. We will refer to $\ell^a$ as the {\em outgoing} null vector.
Third, a stability condition must be imposed on ${\cal S}$, namely we
demand the marginally trapped surface to be a 
{\em spacetime stably outermost}
in the following sense:

\vspace{0.2cm}
{\bf Definition 1.~} {\em Given a closed marginally trapped surface ${\cal S}$
we will refer to it as spacetime stably outermost
if there exists an outgoing ($-k^a$-oriented) vector $X^a= \gamma \ell^a - 
\psi k^a$, with 
$\gamma\geq0$ and $\psi>0$, such that the variation
of $\theta^{(\ell)}$ with respect to $X^a$ fulfills the condition
\bea
\label{e:stability_condition}
\delta_X \theta^{(\ell)} \geq 0.
\eea
If, in addition, $X^a$ (in particular $\gamma$, $\psi$)
and $\Omega^{(\ell)}_a$ are axisymmetric,
we will refer
to $\delta_X \theta^{(\ell)}\geq 0$ as an (axisymmetry-compatible) 
spacetime stably outermost condition.
}

\vspace{0.2cm}
Here $\delta$ denotes
a  variation operator associated with a deformation of the surface
${\cal S}$ (c.f. for example \cite{BooFai07,AndMarSim05}).
Two remarks are in order. First, note that the characterization 
of a marginally trapped surface
as spacetime stably outermost is independent of the choice
of future-oriented null normals $\ell^a$ and $k^a$. Indeed, 
given $f>0$, for $\ell'^a=f \ell^a$ and  $k'^a=f^{-1}k^a$ we can write
$X^a = \gamma \ell^a - \psi k^a= \gamma' \ell'^a - \psi' k'^a$
(with $\gamma'= f^{-1}\gamma\geq0$ and $\psi'= f \psi>0$), and it holds  
$\delta_X \theta^{(\ell')}=f \cdot \delta_X \theta^{(\ell)}>0$.
Second, the proof of inequality (\ref{e:inequality}) would only require
the vector $X^a$ in the stability condition to be outgoing past null
\cite{Hay94,Racz:2008tf}. 
We have, however, kept a more generic characterization in Definition 1
that directly extends the stably outermost condition in \cite{AndMarSim05}
(in particular, ${\cal S}$ is spacetime stably outermost if there exists 
a ($-k^a$-oriented) vector for which  ${\cal S}$ is stably outermost in the sense
of  \cite{AndMarSim08}).

We can now establish the lower bound on the horizon area
by following analogous steps to those in \cite{Dain:2011pi}.
First, we derive a generic inequality on ${\cal S}$, provided by
the following lemma.

\vspace{0.2cm}
{\bf Lemma 1.~} {\em Given a closed marginally trapped 
 surface ${\cal S}$ satisfying the 
spacetime stably outermost condition for an axisymmetric
$X^a$, then for all axisymmetric $\alpha$ it holds
\bea
\label{e:inequality_alpha}
&&\int_{\cal S} \left[D_a\alpha D^a\alpha + \frac{1}{2} \alpha^2 
\; {}^2\!R \right] dS \geq
 \\ 
&& \int_{\cal S} \left[ \alpha^2 \Omega^{(\eta)}_a  {\Omega^{(\eta)}}^a +
\alpha \beta \sigma^{(\ell)}_{ab} {\sigma^{(\ell)}}^{ab} 
+ G_{ab}\alpha\ell^a (\alpha k^b + \beta\ell^b) \right] dS \nn \ ,
\eea
where $\beta=\alpha\gamma/\psi$.
}

\vspace{0.2cm}
To prove it we basically 
follow the discussion in section 3.3. of 
\cite{Andersson:2010jv}, allowing one to essentially reduce the 
nontime-symmetric case to the time-symmetric one
(cf. Th. 2.1 in \cite{Galloway:2005mf} for a similar reasoning). 
First, we evaluate $\delta_X \theta^{(\ell)}/\psi$  for the vector 
$X^a=\gamma \ell^a - \psi k^a$ provided by Definition 1,
with axisymmetric $\gamma$ and $\psi$
(use e.g. Eqs. (2.23) and (2.24) in \cite{BooFai07})
and impose $\theta^{(\ell)}=0$. We can write 
\bea
\label{e:delta_X_theta}
\frac{1}{\psi}\delta_X\theta^{(\ell)}&=& 
- \frac{\gamma}{\psi} \left[\sigma^{(\ell)}_{ab} {\sigma^{(\ell)}}^{ab} 
+ G_{ab}\ell^a\ell^b \right] \nn \\
&&- {}^2\!\Delta \mathrm{ln}\psi -   
D_a\mathrm{ln}\psi D^a\mathrm{ln}\psi
+ 2 \Omega^{(\ell)}_a  D^a\mathrm{ln}\psi  \\
&&-\left[-D^a  \Omega^{(\ell)}_a 
+ \Omega^{(\ell)}_c  {\Omega^{(\ell)}}^c -\frac{1}{2}{}^2\!R + G_{ab}k^a\ell^b  \right] \nn \ .
\eea
We multiply now the expression by $\alpha^2$ and integrate on ${\cal S}$. Using 
$\int_{\cal S} \frac{\alpha^2}{\psi} \delta_X\theta^{(\ell)}dS\geq 0$,
integrating by parts to remove boundary terms, we can write 
\bea
\label{e:int_delta_X_theta_1}
0\leq && \int_{\cal S}\alpha\beta \left[-\sigma^{(\ell)}_{ab} {\sigma^{(\ell)}}^{ab} 
- G_{ab}\ell^a\ell^b \right] dS\nn \\
&+&\int_{\cal S} \alpha^2 \left[-\Omega^{(\ell)}_a  {\Omega^{(\ell)}}^a 
+\frac{1}{2}{}^2\!R - G_{ab}k^a\ell^b \right] dS \nn \\ 
&+& \int_{\cal S}\left[2\alpha D_a\alpha D^a \mathrm{ln}\psi -   
\alpha^2 D_a\mathrm{ln}\psi D^a\mathrm{ln}\psi\right]
dS 
\nn \\
&+& \int_{\cal S}\left[2\alpha^2\Omega^{(\ell)}_a D^a\mathrm{ln}\psi 
- 2\alpha \Omega^{(\ell)}_a D^a\alpha\right]dS \ .
\eea
From the axisymmetry of $\alpha$ and $\psi$, ${\Omega^{(\eta)}}^aD_a\alpha=
{\Omega^{(\eta)}}^aD_a\psi=0$, and using (\ref{e:Omega_eta_xi})
we can write
\bea
\label{e:int_delta_X_theta_2}
0\leq && \int_{\cal S}\alpha\beta \left[-\sigma^{(\ell)}_{ab} {\sigma^{(\ell)}}^{ab} 
- G_{ab}\ell^a\ell^b \right] dS\nn \\
&+&\int_{\cal S} \alpha^2 \left[-\Omega^{(\eta)}_a  {\Omega^{(\eta)}}^a 
+\frac{1}{2}{}^2\!R - G_{ab}k^a\ell^b \right] dS \nn \\ 
&+& 
\int_{\cal S}\left[2 (D^a\alpha) (\alpha D_a \mathrm{ln}\psi -\alpha \Omega^{(\xi)}_a) 
\right. \\
&&\left.-(\alpha D_a\mathrm{ln}\psi -  \alpha \Omega^{(\xi)}_a)
(\alpha D^a\mathrm{ln}\psi -  \alpha {\Omega^{(\xi)}}^a)
\right]dS \ . \nn
\eea
Making use of the Young's inequality in the last integral
\bea
\label{e:Young}
D^a\alpha D_a\alpha \geq 2 D^a\alpha (\alpha D_a \mathrm{ln}\psi -\alpha \Omega^{(\xi)}_a)
-|\alpha D\mathrm{ln}\psi -  \alpha \Omega^{(\xi)}|^2\nn
\eea
inequality (\ref{e:inequality_alpha}) follows for all axisymmetric $\alpha$.

Inequality  (\ref{e:inequality_alpha})  constitutes the first key ingredient in 
the present discussion and the counterpart of inequality
(15)  in \cite{Dain:2011pi} [inserting their Eqs. (30) and (31)].
In this spacetime version, the geometric meaning of each term in inequality 
(\ref{e:inequality_alpha}) is apparent. 
For our present purposes, we first disregard the positive-definite gravitational radiation 
shear squared term. Imposing Einstein equations, we also disregard 
the cosmological constant and matter terms \footnote{In the study of the 
charged case, the stress-energy tensor of the electromagnetic
field is kept. In this case, expression (\ref{e:angular_momentum}) must
be completed with an angular momentum electromagnetic contribution
\cite{AshBeeLew01}. 
This will be addressed in a forthcoming work.}, 
under the assumption of non-negative cosmological constant $\Lambda\geq 0$
and the dominant energy condition
(note that $\alpha k^b + \beta\ell^b$ is a non-spacelike vector).
Therefore
\be
\label{e:geom_inequality}
\int_{\cal S} \left[D_a\alpha D^a\alpha + \frac{1}{2} \alpha^2 
\; {}^2\!R \right] dS\geq \int_{\cal S} \alpha^2 \Omega^{(\eta)}_a  {\Omega^{(\eta)}}^a dS.
\ee
This geometric inequality completes the first stage towards the lower bound
on $A$.

In a second stage, under the assumption of axisymmetry we evaluate 
inequality (\ref{e:geom_inequality})
along the lines in \cite{Dain:2011pi}. First, we note that
the sphericity of ${\cal S}$ follows from
Lemma 1 under the outermost stably and dominant energy conditions 
together with  $\Lambda\geq 0$
since, upon the choice of a constant 
$\alpha$ in Eq. (\ref{e:inequality_alpha}), it implies (for non-vanishing angular momentum)
a positive value for the Euler characteristic of ${\cal S}$. Then,
the following form for the axisymmetric  
line element on ${\cal S}$ is adopted 
\bea
\label{e:q_ab}
ds^2=q_{ab}dx^a dx^b = e^\sigma \left(e^{2q} d\theta^2 + 
\mathrm{sin}^2\theta d\varphi^2 \right) \ ,
\eea
with $\sigma$ and $q$  functions on $\theta$ satisfying $\sigma+q=c$, where $c$ is a constant. 
This coordinate system 
can always be found in axisymmetry \footnote{Note that 
this is essentially the coordinate
system employed in the definition of isolated horizon mass multipoles
in \cite{AshEngPaw04}.}. 
We can then write $dS=e^c dS_0$, with 
$dS_0= \mathrm{sin}\theta d\theta d\varphi$. In addition,
the squared norm $\eta$ of the axial Killing vector $\eta^a=(\partial_\varphi)^a$ is given by $\eta = e^\sigma \mathrm{sin}^2\theta$. 

Regarding the left hand side in (\ref{e:geom_inequality}), we proceed exactly 
as in \cite{Dain:2011pi}. In particular, 
choosing $\alpha=e^{c-\sigma/2}$, the evaluation of
the left-hand-side in inequality (\ref{e:geom_inequality}) 
results in (see \cite{Dain:2011pi})
\bea
\label{e:analytic_lhs}
&&\int_{\cal S} \left[D_a\alpha D^a\alpha + \frac{1}{2} \alpha^2 
\; {}^2\!R \right] dS \\
&&= e^c\left[4\pi(c+1)-\int_{\cal S}\left(\sigma+\frac{1}{4}
\left(\frac{d\sigma}{d\theta}\right)^2\right)
dS_0\right]. \nn
\eea
The second key ingredient in the present discussion concerns the 
evaluation of the right hand side in (\ref{e:geom_inequality}), in particular
the possibility of making contact with the variational functional 
${\cal M}$ employed in  
\cite{Acena:2010ws,Dain:2011pi}. 

Due to the $S^2$ topology of ${\cal S}$, we can always express
$\Omega^{(\ell)}_a$ in terms of a divergence-free and an exact
form. Writing
\bea
\label{e:Omega_ell}
\Omega^{(\ell)}_a = \frac{1}{2\eta} \epsilon_{ab}D^b \bar{\omega}
+D_a\lambda \ ,
\eea
with $\bar{\omega}$ and $\lambda$ fixed up to a constant,
from the axisymmetry of $q_{ab}$ and $\Omega^{(\ell)}_a$ (functions
$\bar{\omega}$ and $\lambda$ are then axially symmetric) it follows that
$\Omega^{(\eta)}_a=  \frac{1}{2\eta} \epsilon_{ab}D^b \bar{\omega}$
is the divergence-free part whereas $\Omega^{(\xi)}_a$
is the exact (gauge) part. In particular,
$\eta^a \Omega^{(\ell)}_a =  \frac{1}{2\eta} \epsilon_{ab}\eta^a 
D^b\bar \omega$ and expressing $\xi^a$
as $\xi_b=\eta^{-1/2} \epsilon_{ab}\eta^a$, we have
\be
\label{e:eta_Omega}
   \Omega^{(\ell)}_a  \eta^a=  \frac{1}{2\eta^{1/2}}\xi^a D_a\bar \omega \ .
\ee
Plugging this expression into Eq. (\ref{e:angular_momentum}) and using 
(\ref{e:q_ab}) we find
\be
\label{eq:J_omega}
  J=\frac{1}{8}\int_{0}^\pi  \partial_\theta\bar \omega \, d\theta
   =\frac{1}{8} \left(\bar \omega(\pi)-\bar \omega(0) \right) 
\ee
which is identical to the relation between $J$ and the {\em twist 
potential} $\omega$ in Eq. (12) of \cite{Acena:2010ws}. As a 
remark, we note that if the axial vector $\eta^a$ on ${\cal S}$ extends 
to a spacetime neighbourhood of ${\cal S}$ 
(something not needed in the present discussion),
we can define the {\em twist} vector of $\eta^a$ as
$\omega_a= \epsilon_{abcd}\eta^b \nabla^c\eta^d$ and the relation
$\xi^a \omega_a=\xi^aD_a\bar{\omega}$ holds. In the vacuum case,
a twist potential $\omega$ satisfying $\omega_a = \nabla_a\omega$
can be defined, so that $\bar{\omega}$ and $\omega$ coincide on
${\cal S}$ up to a
constant. 
Note however that $\bar{\omega}$ on ${\cal S}$ can be defined always.

From Eqs. (\ref{e:Omega_ell}) and (\ref{e:q_ab}) and the choice of 
$\alpha$, we have
\be
\label{e:rhs_ineq}
\alpha^2 \Omega^{(\eta)}_a  {\Omega^{(\eta)}}^a 
=\frac{\alpha^2}{4\eta^2} D_a\bar{\omega} D^a\bar{\omega} = 
\frac{1}{4\eta^{2}}\left(\frac{d\bar{\omega}}{d\theta}\right)^{2} \ .
\ee
Using this and (\ref{e:analytic_lhs}) in (\ref{e:geom_inequality}) 
we recover {\em exactly} the bound 
\bea
\label{e:area_bound}
A\geq 4\pi e^{\frac{{\cal M} -8}{8}} \ \ ,
\eea
with the action functional
\bea
\label{e:functional}
{\cal M}= \frac{1}{2\pi} \int_{\cal S}
\left[\left(\frac{d\sigma}{d\theta}\right)^2 + 
4\sigma
+ \frac{1}{\eta^2} \left(\frac{d\bar{\omega}}{d\theta}\right)^2
\right]dS_0 \ ,
\eea
in Ref. \cite{Dain:2011pi}, so that the rest of the proof reduces to that 
in this reference. Namely, the upper bound in  \cite{Acena:2010ws}
for $J$ 
\bea
\label{e:J_bound}
e^{({\cal M} -8)/8}\geq 2|J| \ ,
\eea
together with inequality (\ref{e:area_bound}) lead to
the area-angular momentum inequality (\ref{e:inequality}) and, in addition, 
a rigidity result follows: if equality in 
(\ref{e:inequality}) holds, first, the intrinsic geometry of ${\cal S}$
is that of an extreme Kerr throat sphere \cite{Dain:2010qr}
and, second, the vanishing of the positive-definite terms 
in (\ref{e:inequality_alpha}) implies in particular,  
for spacelike $X^a$ in (\ref{e:stability_condition}),
the vanishing of the shear $\sigma^{(\ell)}_{ab}$
so that ${\cal S}$ is an {\em instantaneous }
(non-expanding) isolated horizon \cite{AshKri04}.

\emph{Discussion.~}
We have shown that axisymmetric stable marginally trapped surfaces 
(in particular, 
apparent horizons)
satisfy the inequality $A\geq 8\pi|J|$
in generically dynamical, non-necessarily axisymmetric, spacetimes with 
ordinary matter that can extend to the horizon.
There are two key ingredients enabling the remarkable shift from 
the initial data discussion
of inequality (\ref{e:inequality}) in \cite{Dain:2011pi} to a (purely quasi-local)
spacetime result. First,
the derivation of the  geometric inequality (\ref{e:inequality_alpha}) 
where the spacetime interpretation of each term in the right hand side is transparent 
and, more importantly, the global sign is controlled by standard physical assumptions 
on the matter energy content.
This relaxes the counterpart maximal slicing hypothesis in \cite{Dain:2011pi}.
Second, using the spherical topology of ${\cal S}$ we
express the quadratic term controlling the angular momentum 
in the inequality in terms of a potential $\bar{\omega}$
living solely on 
the sphere and leading to an exact match with the key variational 
functional in \cite{Dain:2011pi}. 
This permits to avoid any further assumption on the spacetime geometry.
A critical ingredient in the present derivation is the stability assumption
(\ref{e:stability_condition}), basically the {\em stably outermost} condition
in \cite{AndMarSim08} that naturally extends
the stability condition in \cite{AndMarSim05} (in the context of
spatial 3-slices) to general spacetime embeddings of 
marginally trapped surfaces. 
This stability condition, essentially equivalent
to the {\em outer} horizon condition in \cite{Hay94},
implies that (axially symmetric)\footnote{The non-axially symmetric case 
is of astrophysical interest. Ambiguities in the general
definition of $J$, in particular involving the choice
and normalization of the axial vector $\eta^a$, complicate the writing
of a universal area-angular momentum inequality. The present analysis however
suggests that such upper bounds on appropriate $J$'s are
to be expected, in particular when close to axisymmetry, this being the subject
of current research. 
} 
outer trapping horizons \cite{Hay94}
satisfy inequality (\ref{e:inequality}), independently
of a future/past condition 
on $\theta^{(k)}$ (respectively, $\theta^{(k)}<0$ or $\theta^{(k)}>0$). 
Therefore, quasi-local models
of black hole horizons \cite{Hay94,AshKri04,BooFai07}
satisfy the area-angular momentum inequality (\ref{e:inequality}),
that provides a quasi-local 
characterization of black hole (sub)extremality. 
In particular, 
the validity of (\ref{e:inequality}) is equivalent to the 
non-negativity of the surface gravity $\kappa$
of dynamical and isolated horizons \cite{AshKri04}
(with $\kappa=0$ in the extremal case), the present result therefore
endorsing the physical consistency of their associated first law of
black hole thermodynamics \cite{AshKri02}.
Finally, in Ref. \cite{Dain:2011pi} the following question is posed: 
{\em how small a black hole can be?} Though, according 
to  inequality (\ref{e:inequality})  rotating classical
black holes cannot be arbitrarily small, under the light
of Eq. (\ref{e:inequality_alpha}) one could have violations
of $A\geq 8 \pi |J|$ in near extremal semi-classical collapse
due to corrections violating the dominant energy condition, in particular
relevant when the black hole is small. This is also consistent
with the violations of inequality (\ref{e:inequality}) found in
Ref. \cite{Bode:2011xz}, in the context of black holes accreting matter that
violates the null (and therefore the dominant) energy condition.
Equation (\ref{e:inequality_alpha})
provides a tool to estimate such possible violations.


\noindent\emph{Acknowledgments.~} 
It is a pleasure for J.L.J. to thank M. Ansorg for highlighting 
the physical and geometrical relevance of this problem and for his 
encouragement. S.D. thanks J.M. Espinar for discussions. 
S.D. and J.L.J. thank A. Ace\~na, M.E. Gabach Cl\'ement
for enlightening discussions.
J.L.J. also thanks C. Barcel\'o, R.P. Macedo, M. Mars, 
A. Nielsen and J.M.M. Senovilla. 
J.L.J. acknowledges
the Alexander von Humboldt Foundation, the Spanish MICINN 
(FIS2008-06078-C03-01) and
the Junta de Andaluc\'\i a (FQM2288/219).
S.D. is supported by CONICET (Argentina). This work was
supported by grant PIP 6354/05 of CONICET (Argentina), grant Secyt-UNC
(Argentina) and the Partner Group grant of the MPI for
Gravitational Physics (Germany).






\end{document}